\documentclass[12 pt,twoside,aps,prd,amsmath,amssymb,
tightenlines,showpacs,showkeys,eqsecnum]{revtex4}
\usepackage{mathptmx}
\usepackage{bm}
\usepackage{graphicx}
\usepackage{hyperref}
\renewcommand{\cal}{\mathcal}

\newcommand {\ve}{\varepsilon}
\newcommand {\pr}{\partial}
\newcommand {\prm}{\prime}
\newcommand {\cG}{\cal G}
\newcommand {\cL}{\cal L}
\newcommand {\cD}{\cal D}
\newcommand {\G}{\Gamma}
\newcommand {\bg}{\bar \gamma}
\newcommand {\bp}{\bar \psi}

\newcommand {\p}{\psi}
\newcommand {\vf}{\varphi}

\def \myfigures #1#2#3#4#5#6#7#8
{\begin{figure}[ht]
    \begin{center}
        \includegraphics[width=#2 \textwidth]{#1.eps}
        \hfill
        \includegraphics[width=#6 \textwidth]{#5.eps}
        \parbox[t]{#4\textwidth}{\caption {#3}\label{#1}}
        \hfill
        \parbox[t]{#8\textwidth}{\caption {#7}\label{#5}}
    \end{center}
\end{figure} }

\def\myfigure #1#2#3#4
{\begin{figure}[ht]\begin{center}
\includegraphics[width=#2 \textwidth]{#1.eps}
\parbox[t]{#4\textwidth}{\caption{#3}\label{#1}}
\end{center}\end{figure}}

\date{\today}
\begin{document}
\title{Interacting spinor and scalar fields in Bianchi type-I Universe filled
with viscous fluid: exact and numerical solutions}
\author{Bijan Saha}
\affiliation{Laboratory of Information Technologies\\
Joint Institute for Nuclear Research, Dubna\\
141980 Dubna, Moscow region, Russia} \email{saha@thsun1.jinr.ru,
bijan@jinr.ru} \homepage{http://thsun1.jinr.ru/~saha/}

\begin{abstract}
We consider a self-consistent system of spinor and scalar fields
within the framework of a Bianchi type I gravitational field filled
with viscous fluid in presence of a $\Lambda$ term. Exact self-consistent
solutions to the corresponding spinor, scalar and BI gravitational field
equations are obtained in terms of $\tau$, where $\tau$ is the volume scale
of BI universe. System of equations for $\tau$ and $\ve$, where $\ve$ is the
energy of the viscous fluid, is deduced. Some special cases allowing exact
solutions are thoroughly studied.
\end{abstract}



\maketitle

\bigskip


\section{Introduction}

The investigation of relativistic cosmological models usually has
the energy momentum tensor of matter generated by a perfect fluid.
To consider more realistic models one must take into account the
viscosity mechanisms, which have already attracted the attention
of many researchers
\cite{mis1,mis2,wein,weinb,lang,waga,pacher,guth,murphy,santos}.

The nature of cosmological solutions for homogeneous Bianchi type
I (BI) model was investigated by Belinsky and Khalatnikov
\cite{belin} by taking into account dissipative process due to
viscosity. In \cite{Visprd04,Visrykh04} we reinvestigate the
problem posed in \cite{belin} in presence of a $\Lambda$ term.
Though that Murphy \cite{murphy} claimed that the introduction of
bulk viscosity can avoid the initial singularity at finite past,
but Belinsky and Khalatnikov \cite{belin} showed that viscosity
cannot remove the cosmological singularity but results in a
qualitatively new behavior of the solutions near singularity. To
eliminate the initial singularities a self-consistent system of
nonlinear spinor and BI gravitational field was considered by us
in a series of papers \cite{sahajmp,sahaprd,sahagrg,sahal}. For
some cases we were able to find field configurations those were
always regular. Recently it was found that the introduction of a
spinor field into the system may explain the late time
acceleration of the Universe, hence can be considered as an
alternative to dark energy \cite{SpinDErevprd}.

Given the importance of viscous fluid and spinor field to
construct a more realistic model of the Universe, we in
\cite{VisRR,grqc0703085} we introduced spinor field into the
system and solved the system for some special choice of viscosity.
The purpose of this paper is to study an interacting system of
spinor and scalar fields within the scope of a Bianchi type I
cosmological model filled with viscous fluid in presence of a
$\Lambda$ term and clarify the role of viscosity and field
interaction in the evolution of the Universe.

        \section{Derivation of Basic Equations}
In this section we derive the fundamental equations for the
interacting spinor, scalar and gravitational fields from the
action and write their solutions in term of the volume scale
$\tau$ defined bellow \eqref{taudef}. We also derive the equation
for $\tau$ which plays the central role here.


We consider a system of nonlinear spinor, scalar and BI gravitational
field in presence of perfect fluid given by the action
\begin{equation}
{\cal S}(g; \p, \bp) = \int\, {\cal L} \sqrt{-g} d\Omega
\label{action}
\end{equation}
with
\begin{equation}
{\cal L} = {\cal L}_{\rm g} + {\cal L}_{\rm ss} + {\cal L}_{\rm
m}. \label{lag}
\end{equation}
The gravitational part of the Lagrangian \eqref{lag} is given by a
Bianchi type I (BI hereafter) space-time, whereas ${\cal L}_{\rm
ss}$ describes the interacting spinor and scalar field lagrangian
and ${\cal L}_{\rm m}$ stands for the lagrangian density of
viscous fluid.

             \subsection{Material field Lagrangian}
We choose the interacting spinor and scalar field Lagrangian as
\begin{equation}
{\cal L}_{ss}=\frac{i}{2} \biggl[ \bp \gamma^{\mu} \nabla_{\mu}
\p- \nabla_{\mu} \bp \gamma^{\mu} \p \biggr] - m\bp \p +
\frac{1}{2}\vf_{,\alpha}\vf^{,\alpha}(1 + \lambda F),
\label{nlspin}
\end{equation}
Here $m$ is the spinor mass, $\lambda$ is the coupling constant
and $F = F(I,J)$ with $I = I_S = S^ 2= (\bar \psi \psi)^2$ and $J = I_P = P^2
= (i \bar \psi \gamma^5 \psi)^2$. We would like to mention that there 
are 5 invariants constructed from bilinear spinor from. Using Fierz
transformation it can be shown that among the five invariants only $I$ 
and $J$ are independent as all other can be expressed by them: $I_V = - I_A =
I_S + I_P$ and $I_Q = 2(I_S - I_P)$ \cite{fierz,kaempffer,takahashi,Ber}. 
Since the the bilinear identities appear to have been given in literature only
partially and the relation between the invariants are given as problem 
[cf. eg. \cite{Ber,peskin}], we work them out in the appendix below.  
Therefore, the choice $F = F(I, J)$, describes the nonlinearity in the most 
general of its form
\cite{sahaprd}. Note that setting $\lambda = 0$ in \eqref{nlspin}
we come to the case with minimal coupling.

             \subsection{The gravitational field}
As a gravitational field we consider the Bianchi type I (BI) cosmological
model. It is the simplest model of anisotropic universe that describes
a homogeneous and spatially flat space-time and if filled with perfect
fluid with the equation of state $p = \zeta \ve, \quad \zeta < 1$, it
eventually evolves into a FRW universe \cite{jacobs,chimento2003}. The isotropy of
present-day universe makes BI model a prime candidate for studying the
possible effects of an anisotropy in the early universe on modern-day
data observations. In view of what has been mentioned above we choose
the gravitational part of the Lagrangian \eqref{lag} in the form
\begin{equation}
{\cal L}_{\rm g} = \frac{R}{2\kappa},
\label{lgrav}
\end{equation}
where $R$ is the scalar curvature, $\kappa = 8 \pi G$
being the Einstein's gravitational constant. The gravitational field in
our case is given by a Bianchi type I (BI) metric
\begin{equation}
ds^2 = dt^2 - a^2 dx^2 - b^2 dy^2 - c^2 dz^2,
\label{BI}
\end{equation}
with $a,\, b,\, c$ being the functions of time $t$ only. Here the speed of
light is taken to be unity.

  \subsection{Field equations}
Let us now write the field equations corresponding to the action
\eqref{action}.

Variation of \eqref{action} with respect to spinor field $\psi\,(\bp)$
gives spinor field equations
\begin{subequations}
\label{speq}
\begin{eqnarray}
i\gamma^\mu \nabla_\mu \psi - m \psi + {\cD} \psi +
{\cG} i \gamma^5 \psi &=&0, \label{speq1} \\
i \nabla_\mu \bp \gamma^\mu +  m \bp - {\cD} \bp -
{\cG} i \bp \gamma^5 &=& 0, \label{speq2}
\end{eqnarray}
\end{subequations}
where we denote
$$ {\cD} = \frac{\lambda}{2}\vf_{,\alpha}\vf^{,\alpha}\frac{\pr F}{\pr S},
\quad {\cG} = \frac{\lambda}{2}\vf_{,\alpha}\vf^{,\alpha}\frac{\pr
F}{\pr P}.$$

Varying \eqref{action} with respect to scalar field we find
\begin{equation}
\frac{1}{\sqrt{-g}} \frac{\pr}{\pr x^\nu} \Bigl(\sqrt{-g}
g^{\nu\mu} (1 + \lambda F) \vf_{,\mu}\Bigr) = 0. \label{scfe}
\end{equation}

Variation of \eqref{action} with respect to metric tensor
$g_{\mu\nu}$ gives the Einstein's field equation. For the BI
space-time \eqref{BI} on account of the $\Lambda$ term this system
has the form
\begin{subequations}
\label{BID}
\begin{eqnarray}
\frac{\ddot b}{b} +\frac{\ddot c}{c} + \frac{\dot b}{b}\frac{\dot
c}{c}&=&  \kappa T_{1}^{1} +\Lambda,\label{11}\\
\frac{\ddot c}{c} +\frac{\ddot a}{a} + \frac{\dot c}{c}\frac{\dot
a}{a}&=&  \kappa T_{2}^{2} + \Lambda,\label{22}\\
\frac{\ddot a}{a} +\frac{\ddot b}{b} + \frac{\dot a}{a}\frac{\dot
b}{b}&=&  \kappa T_{3}^{3} + \Lambda,\label{33}\\
\frac{\dot a}{a}\frac{\dot b}{b} +\frac{\dot b}{b}\frac{\dot c}{c}
+\frac{\dot c}{c}\frac{\dot a}{a}&=&  \kappa T_{0}^{0} + \Lambda,
\label{00}
\end{eqnarray}
\end{subequations}
where over dot means differentiation with respect to $t$
and $T_{\nu}^{\mu}$ is the energy-momentum tensor
of the material field given by
\begin{eqnarray}
T_{\mu}^{\rho} &=& \frac{i}{4} g^{\rho\nu} \biggl(\bp \gamma_\mu
\nabla_\nu \psi + \bp \gamma_\nu \nabla_\mu \psi - \nabla_\mu \bar
\psi \gamma_\nu \psi - \nabla_\nu \bp \gamma_\mu \psi \biggr)  \label{tem}\\
& & + (1 - \lambda F)  \vf_{,\mu}\vf^{,\rho} - \delta_{\mu}^{\rho}
{\cL} + T_{\mu\,{\rm m}}^{\,\,\,\nu}. \nonumber
\end{eqnarray}
Here $T_{\mu\,{\rm m}}^{\nu}$ is the energy-momentum tensor of a
viscous fluid having the form
\begin{equation}
T_{\mu\,{\rm m}}^{\nu} = (\ve + p^\prm) u_\mu u^\nu - p^{\prm}
\delta_\mu^\nu + \eta g^{\nu \beta} [u_{\mu;\beta}+u_{\beta:\mu}
-u_\mu u^\alpha u_{\beta;\alpha} - u_\beta u^\alpha
u_{\mu;\alpha}], \label{imper}
\end{equation}
where
\begin{equation}
p^{\prm} = p - (\xi - \frac{2}{3} \eta) u^\mu_{;\mu}.
\label{ppr}
\end{equation}
Here $\ve$ is the energy density, $p$ - pressure, $\eta$ and $\xi$
are the coefficients of shear and bulk viscosity, respectively. In
a comoving system of reference such that $u^\mu = (1,\,0,\,0,\,0)$
we have
\begin{subequations}
\begin{eqnarray}
T_{0\,{\rm m}}^{0} &=& \ve, \\
T_{1\,{\rm m}}^{1} &=& - p^{\prm} + 2 \eta \frac{\dot a}{a}, \\
T_{2\,{\rm m}}^{2} &=& - p^{\prm} + 2 \eta \frac{\dot b}{b}, \\
T_{3\,{\rm m}}^{3} &=& - p^{\prm} + 2 \eta \frac{\dot c}{c}.
\end{eqnarray}
\end{subequations}

In the Eqs. \eqref{speq} and \eqref{tem} $\nabla_\mu$ is the
covariant derivatives acting on a spinor field as
~\cite{zhelnorovich,brill}
\begin{equation}
\label{cvd}
\nabla_\mu \psi = \frac{\partial \psi}{\partial x^\mu} -\G_\mu \psi, \quad
\nabla_\mu \bp = \frac{\partial \bp}{\partial x^\mu} + \bp \G_\mu,
\end{equation}
where $\G_\mu$ are the Fock-Ivanenko spinor connection coefficients
defined by
\begin{equation}
\G_\mu = \frac{1}{4} \gamma^\sigma \Bigl(\G_{\mu \sigma}^{\nu} \gamma_{\nu}
- \partial_{\mu} \gamma_{\sigma}\Bigr).
\label{fock}
\end{equation}
For the metric \eqref{BI} one has the following components
of the spinor connection coefficients
\begin{eqnarray}
\G_0 = 0, \quad
\G_1 = \frac{1}{2}\dot a(t) \bg^1 \bg^0, \quad
\G_2 = \frac{1}{2}\dot b(t) \bg^2 \bg^0, \quad
\G_3 = \frac{1}{2}\dot c(t) \bg^3 \bg^0.
\label{ficc}
\end{eqnarray}
The Dirac matrices $\gamma^\mu(x)$ of curved space-time are
connected with those of Minkowski one as follows:
$$ \gamma^0=\bg^0,\quad \gamma^1 =\bg^1/a,
\quad \gamma^2=\bg^2 /b,\quad \gamma^3 =\bg^3 /c$$
with
\begin{eqnarray}
\bg^0\,=\,\left(\begin{array}{cc}I&0\\0&-I\end{array}\right), \quad
\bg^i\,=\,\left(\begin{array}{cc}0&\sigma^i\\
-\sigma^i&0\end{array}\right), \quad
\gamma^5 = \bg^5&=&\left(\begin{array}{cc}0&-I\\
-I&0\end{array}\right),\nonumber
\end{eqnarray}
where $\sigma_i$ are the Pauli matrices:
\begin{eqnarray}
\sigma^1\,=\,\left(\begin{array}{cc}0&1\\1&0\end{array}\right),
\quad
\sigma^2\,=\,\left(\begin{array}{cc}0&-i\\i&0\end{array}\right),
\quad
\sigma^3\,=\,\left(\begin{array}{cc}1&0\\0&-1\end{array}\right).
\nonumber
\end{eqnarray}
Note that the $\bg$ and the $\sigma$ matrices obey the following
properties:
\begin{eqnarray}
\bg^i \bg^j + \bg^j \bg^i = 2 \eta^{ij},\quad i,j = 0,1,2,3
\nonumber\\
\bg^i \bg^5 + \bg^5 \bg^i = 0, \quad (\bg^5)^2 = I,
\quad i=0,1,2,3 \nonumber\\
\sigma^j \sigma^k = \delta_{jk} + i \varepsilon_{jkl} \sigma^l,
\quad j,k,l = 1,2,3 \nonumber
\end{eqnarray}
where $\eta_{ij} = \{1,-1,-1,-1\}$ is the diagonal matrix,
$\delta_{jk}$ is the Kronekar symbol and $\varepsilon_{jkl}$
is the totally antisymmetric matrix with $\varepsilon_{123} = +1$.

We study the space-independent solutions to the spinor and scalar
field equations \eqref{speq} so that $\psi=\psi(t)$ and $\vf =
\vf(t)$. Here we define
\begin{equation}
\tau = a b c = \sqrt{-g}
\label{taudef}
\end{equation}

Under this assumption from \eqref{scfe} for the scalr field we
find
\begin{equation}
\vf = C \int [\tau (1 + \lambda F)]^{-1} dt, \qquad C = {\rm
const.} \label{sfsol}
\end{equation}

The spinor field equation \eqref{speq1} in account of \eqref{cvd}
and \eqref{ficc} takes the form
\begin{equation} i\bg^0
\biggl(\frac{\partial}{\partial t} +\frac{\dot \tau}{2 \tau} \biggr) \psi
- m \psi + {\cD}\psi + {\cG} i \gamma^5 \psi = 0.
\label{spq}
\end{equation}
Setting $V_j(t) = \sqrt{\tau} \psi_j(t), \quad j=1,2,3,4,$ from
\eqref{spq} one deduces the following system of equations:
\begin{subequations}
\label{V}
\begin{eqnarray}
\dot{V}_{1} + i (m - {\cD}) V_{1} - {\cG} V_{3} &=& 0, \\
\dot{V}_{2} + i (m - {\cD}) V_{2} - {\cG} V_{4} &=& 0, \\
\dot{V}_{3} - i (m - {\cD}) V_{3} + {\cG} V_{1} &=& 0, \\
\dot{V}_{4} - i (m - {\cD}) V_{4} + {\cG} V_{2} &=& 0.
\end{eqnarray}
\end{subequations}

From \eqref{speq1} we also write the equations for the invariants
$S,\quad P$ and $A = \bp \bg^5 \bg^0 \psi$
\begin{subequations}
\label{inv}
\begin{eqnarray}
{\dot S_0} - 2 {\cG}\, A_0 &=& 0, \label{S0}\\
{\dot P_0} - 2 (m - {\cD})\, A_0 &=& 0, \label{P0}\\
{\dot A_0} + 2 (m - {\cD})\, P_0 + 2 {\cG} S_0 &=& 0, \label{A0}
\end{eqnarray}
\end{subequations}
where $S_0 = \tau S, \quad P_0 = \tau P$, and $ A_0 = \tau A.$ The
Eq. \eqref{inv} leads to the following relation
\begin{equation}
S^2 + P^2 + A^2 =  C_1^2/ \tau^2, \qquad C_1^2 = {\rm const.}
\label{inv1}
\end{equation}

Giving the concrete form of $F$ from \eqref{V} one writes
the components of the spinor functions in explicitly and
using the solutions obtained one can write the components of
spinor current:
\begin{equation}
j^\mu = \bp \gamma^\mu \psi.
\label{spincur}
\end{equation}
The component $j^0$
\begin{equation}
j^0 = \frac{1}{\tau}
\bigl[V_{1}^{*} V_{1} + V_{2}^{*} V_{2} + V_{3}^{*} V_{3}
+ V_{4}^{*} V_{4}\bigr],
\end{equation}
defines the charge density of spinor field
that has the following chronometric-invariant form
\begin{equation}
\varrho = (j_0\cdot j^0)^{1/2}.
\label{rho}
\end{equation}
The total charge of spinor field is defined as
\begin{equation}
Q = \int\limits_{-\infty}^{\infty} \varrho \sqrt{-^3 g} dx dy dz =
   \varrho \tau {\cal V},
\label{charge}
\end{equation}
where ${\cal V}$ is the volume. From the spin tensor
\begin{equation}
S^{\mu\nu,\epsilon} = \frac{1}{4}\bp \bigl\{\gamma^\epsilon
\sigma^{\mu\nu}+\sigma^{\mu\nu}\gamma^\epsilon\bigr\} \psi.
\label{spin}
\end{equation}
one finds chronometric invariant spin tensor
\begin{equation}
S_{{\rm ch}}^{ij,0} = \bigl(S_{ij,0} S^{ij,0}\bigr)^{1/2},
\label{chij}
\end{equation}
and the projection of the spin vector on $k$ axis
\begin{equation}
S_k = \int\limits_{-\infty}^{\infty} S_{{\rm ch}}^{ij,0}
\sqrt{-^3 g} dx dy dz = S_{{\rm ch}}^{ij,0} \tau V.
\label{proj}
\end{equation}

Let us now solve the Einstein equations. To do it we first write the
expressions for the components of the energy-momentum tensor explicitly:
\begin{subequations}
\label{total}
\begin{eqnarray}
T_{0}^{0} &=& m S + \frac{C^2}{2\tau^2(1+\lambda F)} + \ve \equiv \tilde{T}_{0}^{0},\\
T_{1}^{1} &=& {\cD} S + {\cG} P - \frac{C^2}{2\tau^2(1+\lambda F)}
- p^{\prm} + 2 \eta \frac{\dot a}{a} \equiv \tilde{T}_{1}^{1} + 2 \eta \frac{\dot a}{a}, \\
T_{2}^{2} &=& {\cD} S + {\cG} P - \frac{C^2}{2\tau^2(1+\lambda F)}
- p^{\prm} + 2 \eta \frac{\dot b}{b} \equiv \tilde{T}_{1}^{1} + 2 \eta \frac{\dot b}{b},, \\
T_{3}^{3} &=& {\cD} S + {\cG} P - \frac{C^2}{2\tau^2(1+\lambda F)}
- p^{\prm} + 2 \eta \frac{\dot c}{c} \equiv \tilde{T}_{1}^{1} + 2
\eta \frac{\dot c}{c},.
\end{eqnarray}
\end{subequations}
In account of \eqref{total} from \eqref{BID} we find the metric
functions \cite{sahaprd}
\begin{subequations}
\label{abc}
\begin{eqnarray}
a(t) &=& Y_1 \tau^{1/3}\exp \biggl[\frac{X_1}{3} \int\,\frac{e^{-2
\kappa \int \eta dt}}{\tau (t)} dt \biggr],
\label{a} \\
b(t) &=& Y_2\tau^{1/3}\exp \biggl[\frac{X_2}{3} \int\,\frac{e^{-2
\kappa \int \eta dt} }{\tau (t)} dt \biggr],
\label{b} \\
c(t) &=& Y_3\tau^{1/3}\exp \biggl[\frac{X_3}{3} \int\,\frac{e^{-2
\kappa \int \eta dt}}{\tau (t)} dt \biggr], \label{c}
\end{eqnarray}
\end{subequations}
with the constants $Y_i$ and $X_i$ obeying
$$Y_1Y_2Y_3 = 1, \qquad X_1 + X_2 + X_3 = 0.$$

As one sees from \eqref{a}, \eqref{b} and \eqref{c}, for $\tau = t^n$
with $n > 1$ the exponent tends to unity at large $t$, and the
anisotropic model becomes isotropic one.

Further we will investigate the existence of singularity (singular
point) of the gravitational case, which can be done by
investigating the invariant characteristics of the space-time. In
general relativity these invariants are composed from the
curvature tensor and the metric one. In a 4D Riemann space-time
there are 14 independent invariants. Instead of analyzing all 14
invariants, one can confine this study only in 3, namely the
scalar curvature $I_1 = R$, $I_2 = R_{\mu\nu}^R{\mu\nu}$, and the
Kretschmann scalar $I_3 =
R_{\alpha\beta\mu\nu}R^{\alpha\beta\mu\nu}$. At any regular
space-time point, these three invariants $I_1,\,I_2,\,I_3$ should
be finite. One can easily verify that
$$I_1 \propto \frac{1}{\tau^2},\quad
I_2 \propto \frac{1}{\tau^4},\quad I_3 \propto \frac{1}{\tau^4}.$$
Thus we see that at any space-time point, where $\tau = 0$ the invariants
$I_1,\,I_2,\,I_3$, as well as the scalar and spinor fields
become infinity, hence the space-time becomes singular at this point.

In what follows, we write the equation for $\tau$ and study it in details.

Summation of Einstein equations \eqref{11}, \eqref{22}, \eqref{33} and
\eqref{00} multiplied by 3 gives
\begin{equation}
\ddot \tau = \frac{3}{2}\kappa \Bigl(\tilde{T}_{0}^{0} +
\tilde{T}_{1}^{1}\Bigr)\tau + 3 \kappa \eta \dot \tau + 3 \Lambda
\tau, \label{a4}
\end{equation}
which can be rearranged as
\begin{equation}
{\ddot \tau} - \frac{3}{2} \kappa \xi {\dot \tau} =
\frac{3}{2}\kappa \Bigl(mS + {\cD} S + {\cG} P + \ve - p\Bigr)
\tau + 3 \Lambda \tau. \label{dtau1}
\end{equation}
For the right-hand-side of \eqref{dtau1} to be a function
of $\tau$ only, the solution to this equation is well-known~\cite{kamke}.

On the other hand from Bianchi identity $G_{\mu;\nu}^{\nu} = 0$ one finds
\begin{equation}
T_{\mu;\nu}^{\nu} = T_{\mu,\nu}^{\nu} + \G_{\rho\nu}^{\nu} T_{\mu}^{\rho}
- \G_{\mu\nu}^{\rho} T_{\rho}^{\nu} = 0,
\end{equation}
which in our case has the form
\begin{equation}
\frac{1}{\tau}\bigl(\tau T_0^0\bigr)^{\cdot} - \frac{\dot a}{a} T_1^1
-\frac{\dot b}{b} T_2^2  - \frac{\dot c}{c} T_3^3 = 0.
\label{emcon}
\end{equation}
This equation can be rewritten as
\begin{equation}
\dot{\tilde{T}}_0^0 = \frac{\dot \tau}{\tau}\Bigl(\tilde{T}_{1}^{1} - \tilde{T}_{0}^{0}\Bigr)
+ 2 \eta \Bigl(\frac{\dot a^2}{a^2} + \frac{\dot b^2}{b^2} + \frac{\dot c^2}{c^2}\Bigr).
\label{Biden}
\end{equation}
Recall that \eqref{inv} gives
$$(m -{\cD}) \dot{S}_0 - {\cG} \dot{P}_0 = 0.$$
In view of that after a little manipulation from \eqref{Biden} we obtain
\begin{equation}
{\dot \ve} + \frac{\dot \tau}{\tau} \omega - (\xi + \frac{4}{3}
\eta) \frac{{\dot \tau}^2}{\tau^2} + 4 \eta (\kappa T_0^0 +
\Lambda) = 0, \label{vep}
\end{equation}
where
\begin{equation}
\omega = \ve + p,
\label{thermal}
\end{equation}
is the thermal function. Let us now in analogy with Hubble
constant introduce the quantity $H$, such that
\begin{equation}
\frac{\dot {\tau}}{\tau} = \frac{\dot {a}}{a}+\frac{\dot {b}}{b} +
\frac{\dot {c}}{c} = 3 H.
\label{hubc}
\end{equation}
Then \eqref{dtau1} and \eqref{vep} in account of \eqref{total} can be
rewritten as
\begin{subequations}
\label{HVe}
\begin{eqnarray}
\dot {H} &=& \frac{\kappa}{2}\bigl(3 \xi H - \omega\bigr) -
\bigl(3 H^2 - \kappa \ve  - \Lambda \bigr) + \frac{\kappa}{2}
\bigl(m S + {\cD} S + {\cG} P \bigr),  \label{H}\\
\dot {\ve} &=& 3 H\bigl(3 \xi H - \omega\bigr) + 4 \eta \bigl(3
H^2 - \kappa \ve  - \Lambda  \bigr) - 4 \eta \kappa \Bigl[ m S +
\frac{C^2}{2\tau^2 (1+ \lambda F)}\Bigr]. \label{Ve}
\end{eqnarray}
\end{subequations}
Thus, the metric functions are found explicitly in terms of $\tau$
and viscosity. To write $\tau$ and components of spinor field as
well and scalar one we have to specify the function $F$. In the
next section we explicitly solve Eqs. \eqref{V} and \eqref{HVe}
for some concrete value of $F$.

        \section{Some special solutions}

In this section we first solve the spinor field equations for some special
choice of $F$, which will be given in terms of $\tau$. Thereafter, we
will study the system \eqref{HVe} in details and give explicit solution
for some special cases.

    \subsection{Solutions to the spinor field equations}
As one sees, introduction of viscous fluid has no direct effect on
the system of spinor field equations \eqref{V}. Viscous fluid has an
implicit influence on the system through $\tau$. A detailed analysis
of the system in question can be found in \cite{sahaprd}. Here
we just write the final results.

    \subsubsection{Case with $F = F(I)$}

Here we consider the case when the nonlinear spinor field
is given by
$F =  F(I).$
As in the case with minimal coupling from \eqref{S0} one finds
\begin{equation}
S = \frac{C_0}{\tau}, \quad C_0 = {\rm const.}
\label{stau}
\end{equation}
For  components of spinor field we find~\cite{sahaprd}
\begin{eqnarray}
\psi_1(t) &=& \frac{C_1}{\sqrt{\tau}} e^{-i\beta}, \quad
\psi_2(t) = \frac{C_2}{\sqrt{\tau}} e^{-i\beta},  \nonumber\\
\label{spef}\\
\psi_3(t) &=& \frac{C_3}{\sqrt{\tau}} e^{i\beta}, \quad
\psi_4(t) = \frac{C_4}{\sqrt{\tau}} e^{i\beta},
\nonumber
\end{eqnarray}
with $C_i$ being the integration constants and
are related to $C_0$ as
$C_0 = C_{1}^{2} + C_{2}^{2} - C_{3}^{2} - C_{4}^{2}.$ Here
$\beta = \int(m - {\cD}) dt$.

For the components of the spin current from \eqref{spincur} we find
\begin{eqnarray}
j^0 &=& \frac{1}{\tau} \bigl[C_{1}^{2} + C_{2}^{2} + C_{3}^{2} +
C_{4}^{2}\bigr],\quad j^1 = \frac{2}{a\tau} \bigl[C_{1} C_{4} +
C_{2} C_{3}\bigr] \cos (2\beta),
\nonumber \\
j^2 &=& \frac{2}{b\tau} \bigl[C_{1} C_{4} - C_{2} C_{3}\bigr] \sin
(2\beta),\quad j^3 = \frac{2}{c\tau} \bigl[C_{1} C_{3} - C_{2}
C_{4}\bigr] \cos (2\beta), \nonumber
\end{eqnarray}
whereas, for the projection of spin vectors on the $X$, $Y$ and $Z$
axis we find
\begin{eqnarray}
S^{23,0} = \frac{C_1 C_2 + C_3 C_4}{b c\tau},\quad
S^{31,0} = 0,\quad
S^{12,0} = \frac{C_1^2 - C_2^2 + C_3^2 - C_4^2}{2ab\tau}. \nonumber
\end{eqnarray}
Total charge of the system in a volume $\cal{V}$ in this case is
\begin{equation}
Q = [C_1^2 + C_{2}^{2} + C_{3}^{2} + C_{4}^{2}] \cal{V}.
\end{equation}
Thus, for $\tau \ne 0$ the components of spin current and
the projection of spin vectors are singularity-free and the total
charge of the system in a finite volume is always finite.
Note that, setting $\lambda = 0$, i.e., $\beta = m t$ in the
foregoing expressions one get the results for the linear spinor field.

    \subsubsection{Case with $F = F(J)$}

Here we consider the case with
$F = F(J).$
In this case we assume the spinor field to be massless.
Note that, in the unified
nonlinear spinor theory of Heisenberg, the massive term remains
absent, and according to Heisenberg, the particle mass should be
obtained as a result of quantization of spinor prematter~
\cite{massless}. In the nonlinear generalization of classical field
equations, the massive term does not possess the significance that
it possesses in the linear one, as it by no means defines total
energy (or mass) of the nonlinear field system. Thus without losing
the generality we can consider massless spinor field putting $m\,=\,0.$
Then from \eqref{P0} one gets
\begin{equation}
P = D_0/\tau, \quad D_0 = {\rm const.}
\label{ptau}
\end{equation}
In this case the spinor field components take the form
\begin{eqnarray}
\psi_1 &=&\frac{1}{\sqrt{\tau}} \bigl(D_1 e^{i \sigma} +
iD_3 e^{-i\sigma}\bigr), \quad
\psi_2 =\frac{1}{\sqrt{\tau}} \bigl(D_2 e^{i \sigma} +
iD_4 e^{-i\sigma}\bigr), \nonumber \\
\label{J}\\
\psi_3 &=&\frac{1}{\sqrt{\tau}} \bigl(iD_1 e^{i \sigma} +
D_3 e^{-i \sigma}\bigr),\quad
\psi_4 =\frac{1}{\sqrt{\tau}} \bigl(iD_2 e^{i \sigma} +
D_4 e^{-i\sigma}\bigr). \nonumber
\end{eqnarray}
The integration constants $D_i$
are connected to $D_0$ by
$D_0=2\,(D_{1}^{2} + D_{2}^{2} - D_{3}^{2} -D_{4}^{2}).$
Here we set $\sigma = \int {\cG} dt$.

For the components of the spin current from \eqref{spincur} we find
\begin{eqnarray}
j^0 &=& \frac{2}{\tau}
\bigl[D_{1}^{2} + D_{2}^{2} + D_{3}^{2} + D_{4}^{2}\bigr],\quad
j^1 = \frac{4}{a\tau}
\bigl[D_{2} D_{3} + D_{1} D_{4}\bigr] \cos (2 \sigma), \nonumber\\
j^2 &=& \frac{4}{b\tau} \bigl[D_{2} D_{3} - D_{1} D_{4}\bigr] \sin
(2 \sigma),\quad j^3 = \frac{4}{c\tau} \bigl[D_{1} D_{3} - D_{2}
D_{4}\bigr] \cos (2 \sigma), \nonumber
\end{eqnarray}
whereas, for the projection of spin vectors on the $X$, $Y$ and $Z$
axis we find
\begin{eqnarray}
S^{23,0} = \frac{2(D_{1} D_{2} + D_{3} D_{4})}{b c\tau},\quad
S^{31,0} = 0,\quad
S^{12,0} = \frac{D_{1}^{2} - D_{2}^{2} + D_{3}^{2} - D_{4}^{2}}{2ab\tau}
\nonumber
\end{eqnarray}
We see that for any nontrivial $\tau$ as in previous case
the components of spin current and
the projection of spin vectors are singularity-free and the total
charge of the system in a finite volume is always finite.

 \subsection{Determination of $\tau$}

In this subsection we simultaneously solve the system of equations
for $\tau$ and $\ve$. Since setting $m = 0$ in the equations for
$F= F(I)$ one comes to the case when $F = F(J)$, we consider the
case with $F$ being the function of $I$ only. Let $F$ be the power
function of $S$, i.e., $F = S^n$. As it was established earlier,
in this case $S = C_0/\tau$, or setting $C_0 = 1$ simply $S =
1/\tau$. for simplicity we also set $C = 1$. Evaluating ${\cD}$ in
terms of $\tau$ we then come to the following system of equations
\begin{subequations}
\label{tauve}
\begin{eqnarray}
{\ddot \tau} &=&  \frac{3}{2} \xi {\dot \tau} + \frac{3\kappa}{2}
\Bigl(\frac{m}{\tau} +\frac{\lambda
n}{2}\frac{\tau^{n-1}}{(\lambda + \tau^n)^2}\Bigr)
 + \frac{3\kappa}{2}\Bigl(\ve - p
\Bigr) \tau - 3 \Lambda \tau,\label{dtaun}\\
{\dot \ve} &=& - \frac{\dot \tau}{\tau} \omega + (\xi +
\frac{4}{3} \eta) \frac{{\dot \tau}^2}{\tau^2} - 4 \eta
\Bigl[\frac{m}{\tau} + \frac{\tau^{n-2}}{2(\lambda +\tau^n)} +
\Lambda\Bigr], \label{vepn}
\end{eqnarray}
\end{subequations}
or in terms of $H$
\begin{subequations}
\label{HVen}
\begin{eqnarray}
\dot \tau &=& 3 H \tau, \label{tauH}\\
\dot {H} &=& \frac{1}{2}\bigl(3 \xi H - \omega\bigr) - \bigl(3 H^2
- \ve  - \Lambda\bigr) + \frac{\kappa}{2} \Bigl[ \frac{m}{\tau} +
\frac{\lambda n}{2}\frac{\tau^{n-2}}{(\lambda + \tau^n)^2}\Bigr],
\label{Hn}\\
\dot {\ve} &=& 3 H\bigl(3 \xi H - \omega\bigr) + 4 \eta \bigl(3
H^2 - \ve  - \Lambda \bigr) - 4 \eta \Bigl[ \frac{m}{\tau} +
\frac{\tau^{n-2}}{2(\lambda +\tau^n)} \Bigr]. \label{Ven}
\end{eqnarray}
\end{subequations}
Here $\eta$ and $\xi$ are the bulk and shear viscosity, respectively and
they are both positively definite, i.e.,
\begin{equation}
\eta > 0, \quad \xi > 0.
\end{equation}
They may be either constant or function of time or energy. We consider
the case when
\begin{equation}
\eta = A \ve^{\alpha}, \quad \xi = B \ve^{\beta},
\label{etaxi}
\end{equation}
with $A$ and $B$ being some positive quantities.
For $p$ we set as in perfect fluid,
\begin{equation}
p = \zeta \ve, \quad \zeta \in (0, 1].
\label{pzeta}
\end{equation}
Note that in this case $\zeta \ne 0$, since for dust pressure, hence
temperature is zero, that results in vanishing viscosity.

The system \eqref{HVen} without spinor field have been extensively
studied in literature either partially \cite{murphy,huang,baner}
or as a whole \cite{belin}. Here we try to solve the system
\eqref{tauve} for some particular choice of parameters.

        \subsubsection{Case with bulk viscosity}

Let us first consider the case with bulk viscosity alone setting
coefficient of shear viscosity $\eta = 0$. In this case from
\eqref{a4} and \eqref{Biden} we find the following relation
\begin{equation}
\kappa \tilde{T}_0^0 = 3 H^2 - \Lambda + C_{00}, \quad C_{00} =
{\rm const.} \label{veHrel}
\end{equation}
We also demand the coefficient of bulk viscosity be inverse
proportional to expansion, i.e.,
\begin{equation}
\xi \theta = 3 \xi H = C_2, \quad C_2 = {\rm const.}
\label{bvx}
\end{equation}
Inserting $\eta = 0$, \eqref{bvx} and \eqref{pzeta} into \eqref{Ven} one finds
\begin{equation}
\ve = \frac{1}{1+\zeta} [C_2 - C_3/\tau^{1+\zeta}].\label{vecase1}
\end{equation}
Then from \eqref{dtaun} we get the following equation for
determining  $\tau$:
\begin{equation}
\ddot \tau = \frac{3}{2} \kappa m + 3\Bigl[\frac{C_2}{2}\kappa +
\Lambda\Bigr] \tau + \frac{3\kappa(1 -
\zeta)}{2(1+\zeta)}\frac{C_2 \tau^{1+\zeta} - C_3}{\tau^\zeta} +
\frac{3\kappa \lambda n}{4}\frac{\tau^{n-1}}{(\lambda + \tau^n)^2}
\equiv {\cal F}(q,\tau), \label{dettau1}
\end{equation}
where $q$ is the set of problem parameters. As one sees, the right
hand side of the Eq. \eqref{dettau1} is a function of $\tau$,
hence can be solved in quadrature \cite{kamke}. We solve the Eq.
\eqref{dettau1} numerically. It can be noted that the Eq.
\eqref{dettau1} can be viewed as one describing the motion of a
single particle. Sometimes it is useful to plot the potential of
the corresponding equation which in this case is
\begin{equation}
{\cal U}(q,\tau) = - 2 \int {\cal F}(q,\tau) d\tau. \label{poten1}
\end{equation}
The problem parameters are chosen as follows: $\kappa = 1$, $m =
1$, $\lambda = 0.5$, $\zeta = 1/3$, $n = 4$, $C_2 = 2$ and $C_3 =
1$. Here we consider the cases with different $\Lambda$, namely
with $\Lambda = -2, 0, 1$, respectively. The initial value of
$\tau$ is taken to be a small one, whereas, the first derivative
of $\tau$, i.e., $\dot \tau$ at that point of time is calculated
from \eqref{veHrel}. In Fig. \ref{rfsspoten} we have illustrated
the potential corresponding to Eq. \eqref{dettau1}. It can be
immediately seen that independent of the sign of $\Lambda$ we have
always expanding universe. But as is seen from Fig.
\ref{rfsstauall} a positive $\Lambda$ results in accelerated mode
of expansion, while the negative one causes deceleration.

\myfigures{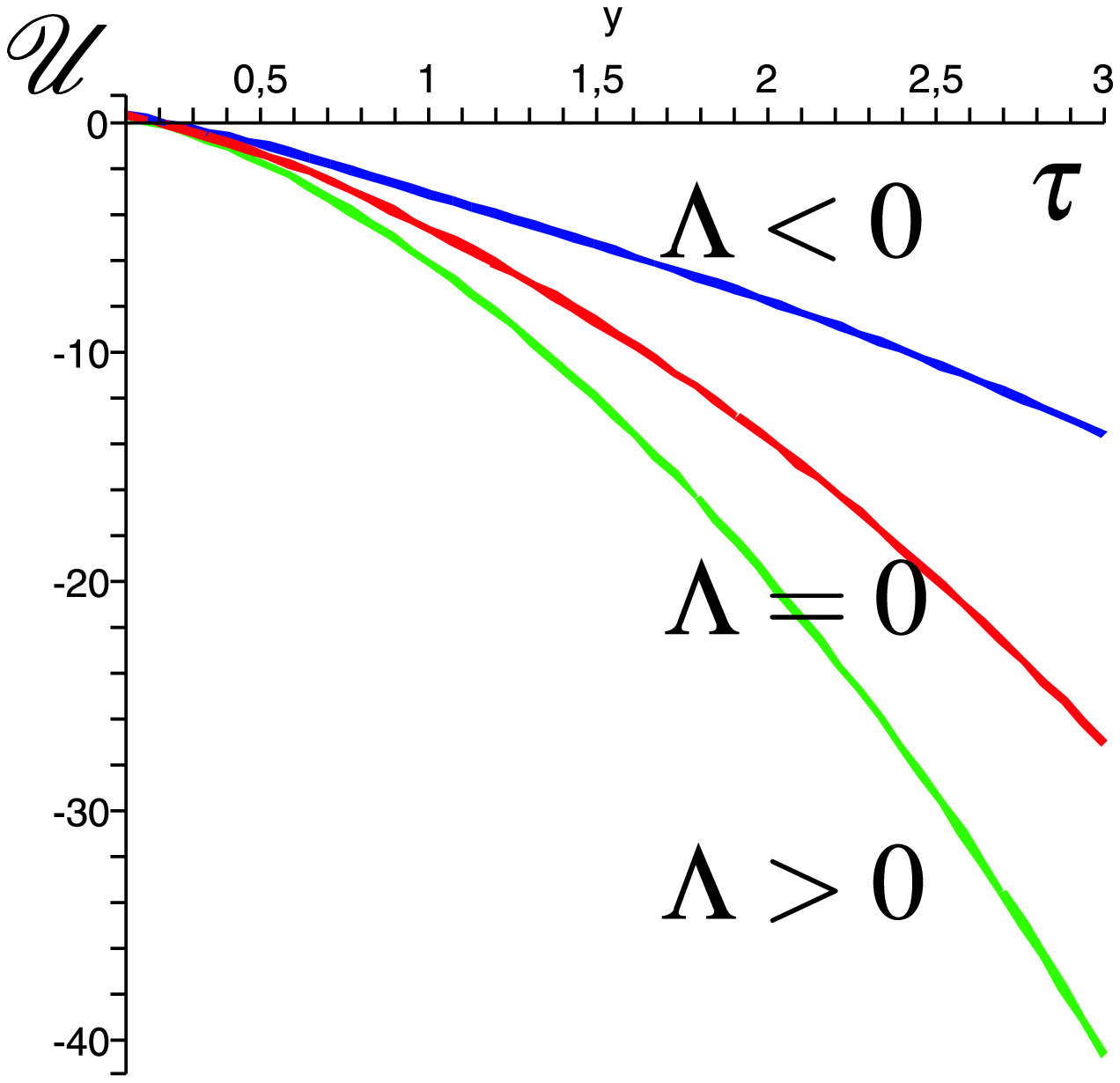}{0.45}{View of the potential corresponding to
the different sign of the $\Lambda$
term.}{0.45}{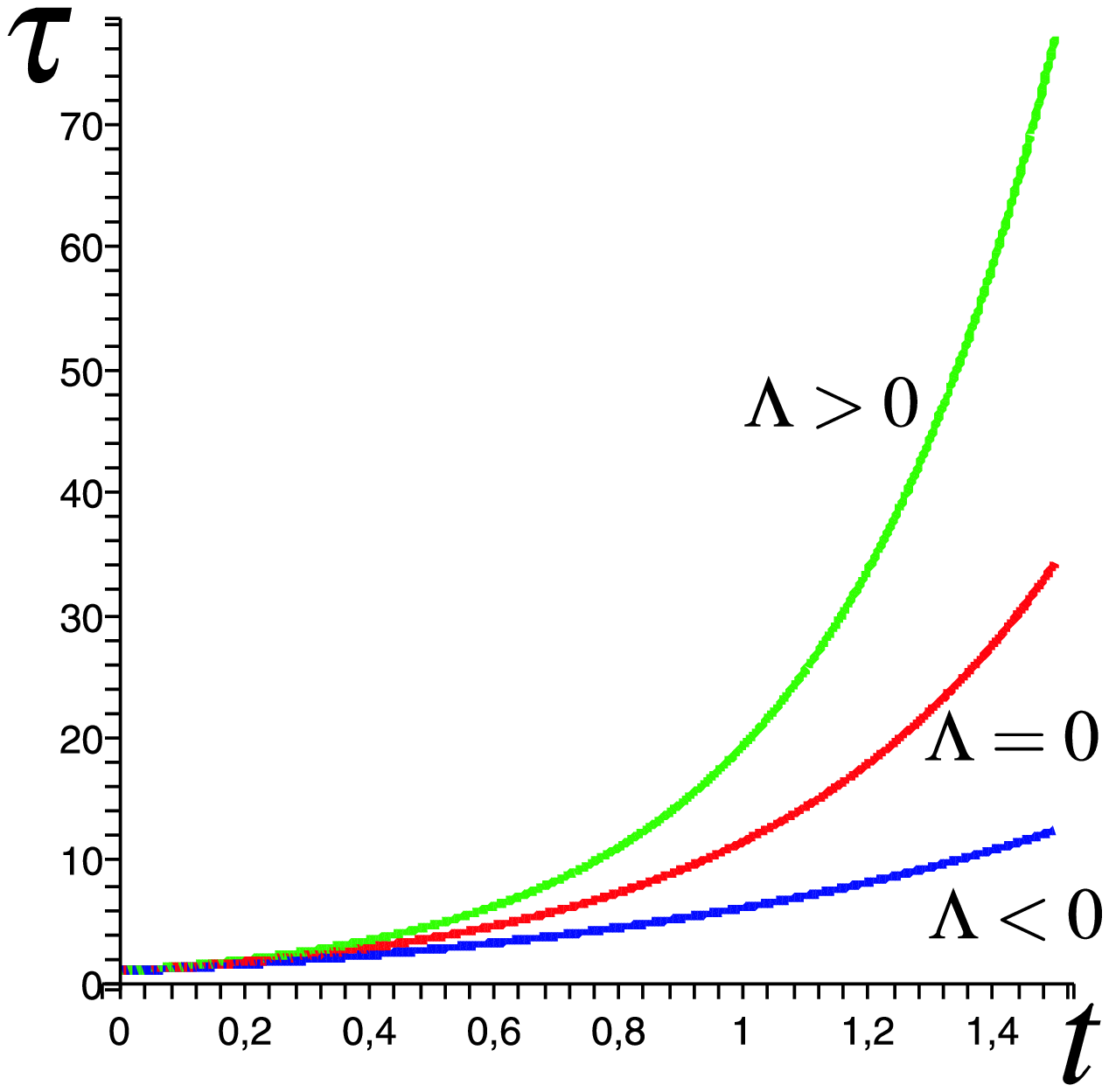}{0.45}{Evolution of $\tau$ depending on
the signs of the $\Lambda$ term.}{0.45}

        \subsubsection{Case with bulk and shear viscosities}
Let us consider more general case. Following \cite{Visprd04} we
choose the shear viscosity being proportional to the expansion,
namely,
\begin{equation}
\eta = - \frac{3}{2\kappa} H = -\frac{1}{2\kappa} \theta.
\label{sbv}
\end{equation}
In absence of spinor field this assumption leads to
\begin{equation}
3H^2 = \kappa \ve + C_4, \quad C_4 = {\rm const.} \label{sbvrel}
\end{equation}
It can be shown that the relation \eqref{sbvrel} in our case can
be achieved only for massless spinor field with the nonlinear term
being
$$ F = (F_0 S^2 - 1)/\lambda.$$
Equation for $\tau$ in this case has the form
\begin{equation}
\tau \ddot \tau - 0.5 (1 - \zeta) \dot \tau^2 - 1.5 \kappa \xi
\tau \dot \tau - 3[\Lambda - 0.5(1 - \zeta)C_4 + \kappa/2F_0]
\tau^2 = 0. \label{sbvtau}
\end{equation}
In case of $\xi = {\rm const.}$ there exists several special
solutions available in handbooks on differential equations. For
that reason we rewrite this equation in terms of $H$:
\begin{equation}
\dot H = - 1.5 (1 + \zeta) H^2 + 1.5 \kappa \xi H + [\Lambda - 0.5
(1 - \zeta) C_4 + \kappa/2F_0]. \label{newH}
\end{equation}
The solution of the foregoing equation can be written in
quadrature as
\begin{equation}
\int\frac{dH}{AH^2 + BH + C} = t, \label{quadr}
\end{equation}
with $A = - 1.5 (1 + \zeta)$, $B = 1.5 \kappa \xi$ and $C =
\Lambda - 0.5 (1 - \zeta) C_4 + \kappa/2F_0$. If the bulk
viscosity is taken to be a constant one, i.e., $\xi = {\rm
const.}$, then depending on the value of the discriminant $B^2 - 4
AC$ there exists three types of solutions, namely \cite{prud}:
\begin{equation}
t = \left\{\begin{array}{lll}\frac{1}{\sqrt{B^2 - 4 A C}} {\rm ln}
|\frac{2 A H + B + \sqrt{B^2 - 4AC}}{2 A H + B - \sqrt{B^2 -
4AC}}|,& & B^2 > 4 A C,\\
& &\\
\frac{2}{\sqrt{4 A C - B^2}} {\rm arctan} \frac{2 A H + B}{\sqrt{4
A C - B^2}}, & & B^2 < 4 A C, \\
& & \\ -\frac{2}{2 A H + B}, & & B^2 = 4 A C. \end{array}\right.
\end{equation}

Note that a detailed analysis of these solutions in absence of
spinor and scalar field was given in \cite{Visprd04}. We choose
the problem parameters as follows: $C_4 = 9$, $\zeta = 1/3$ and
$\kappa = 2$. Under this choice we find $\delta = B^2 - 4AC = 9
\xi^2 + 8(\Lambda - 2)$. After that we chose $\Lambda$ positive,
trivial or negative (in particular we chose $\Lambda = (7/8, 0,
-5/2)$). The quantity $\xi$ now is taken such a way that we have
$\delta = (\delta_1,\delta_2,\delta_3)$ for all values of
$\Lambda$ chosen above, whereas, $\delta_1 > 0$, $\delta_2 = 0$
and $\delta_3 < 0$. In Figs. \ref{delnz}, \ref{delnp} and
\ref{delnm} we plot the evolution of $\tau$ corresponding to a
trivial, positive and negative value of $\delta$, respectively. As
is seen, the behavior of $\tau$ mainly depends on $\delta$ and
independent to the sign of $\Lambda$. Since a negative $\delta$
gives non-periodic mode of evolution we plot the corresponding
phase diagram in Fig. \ref{THm}.

\myfigures{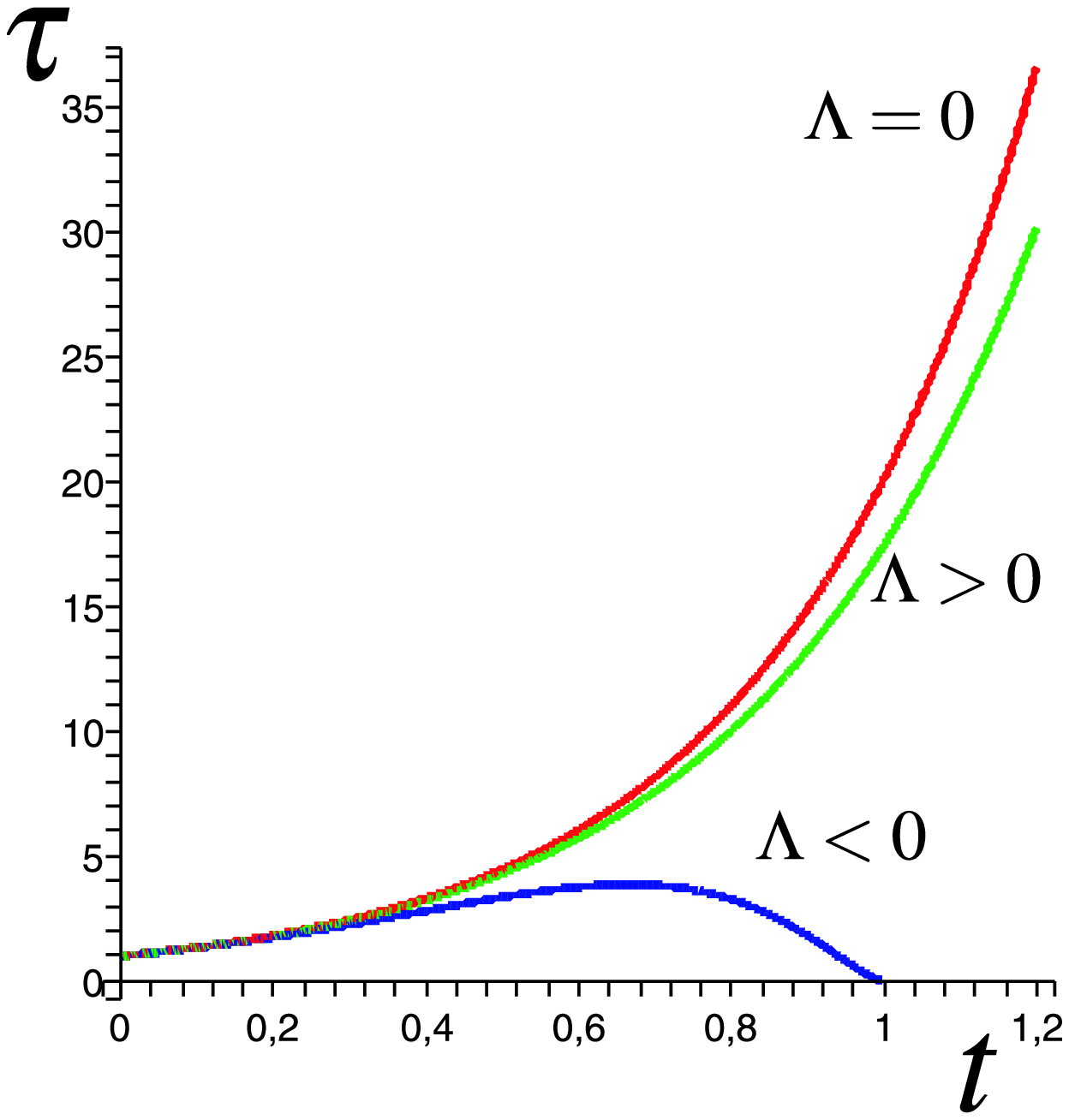}{0.45}{Evolution of the universe with $\delta = 0$ for
different $\Lambda$.}{0.45}{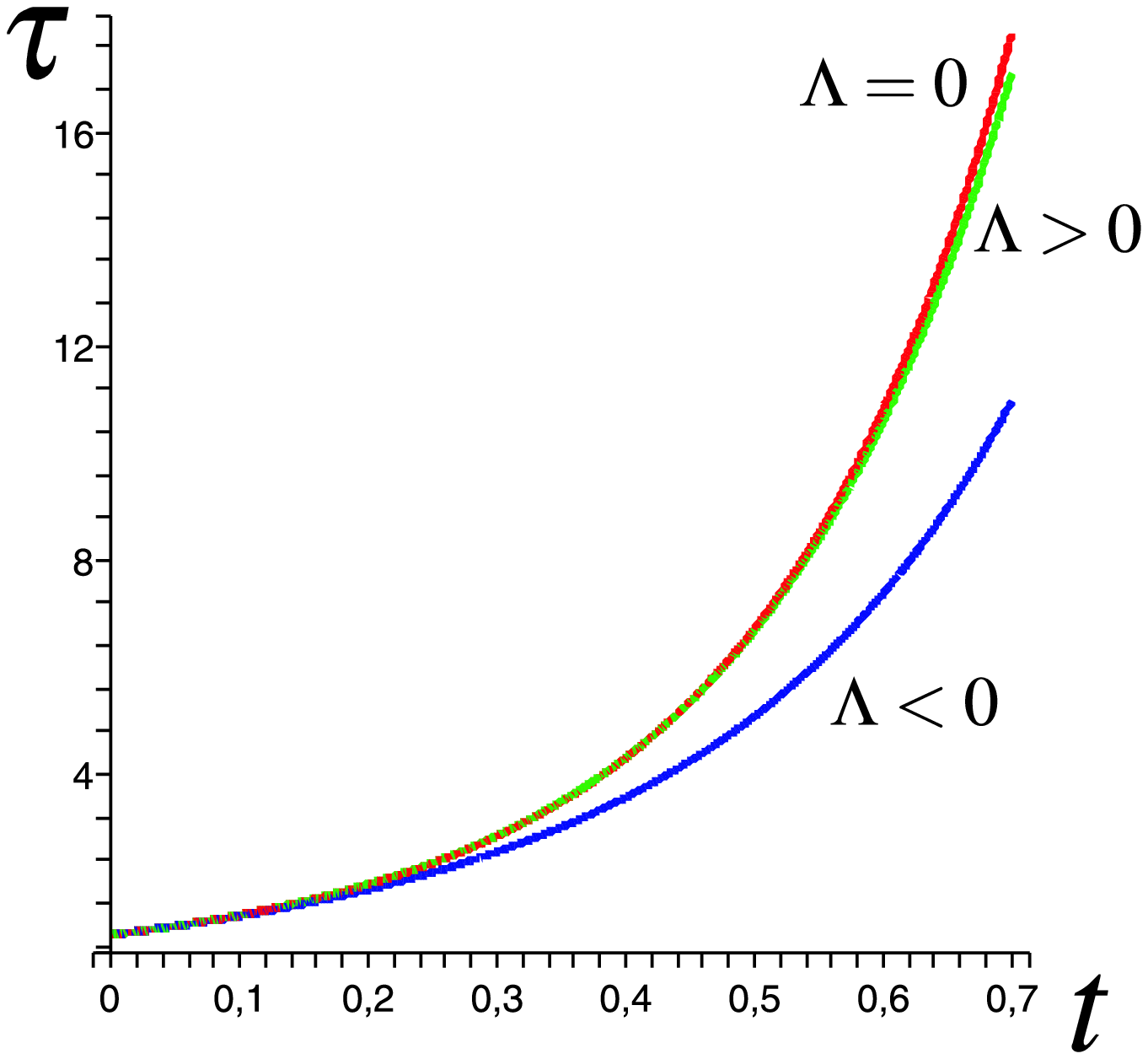}{0.45}{Evolution of the universe
with $\delta > 0$ for different $\Lambda$.}{0.45}

\myfigures{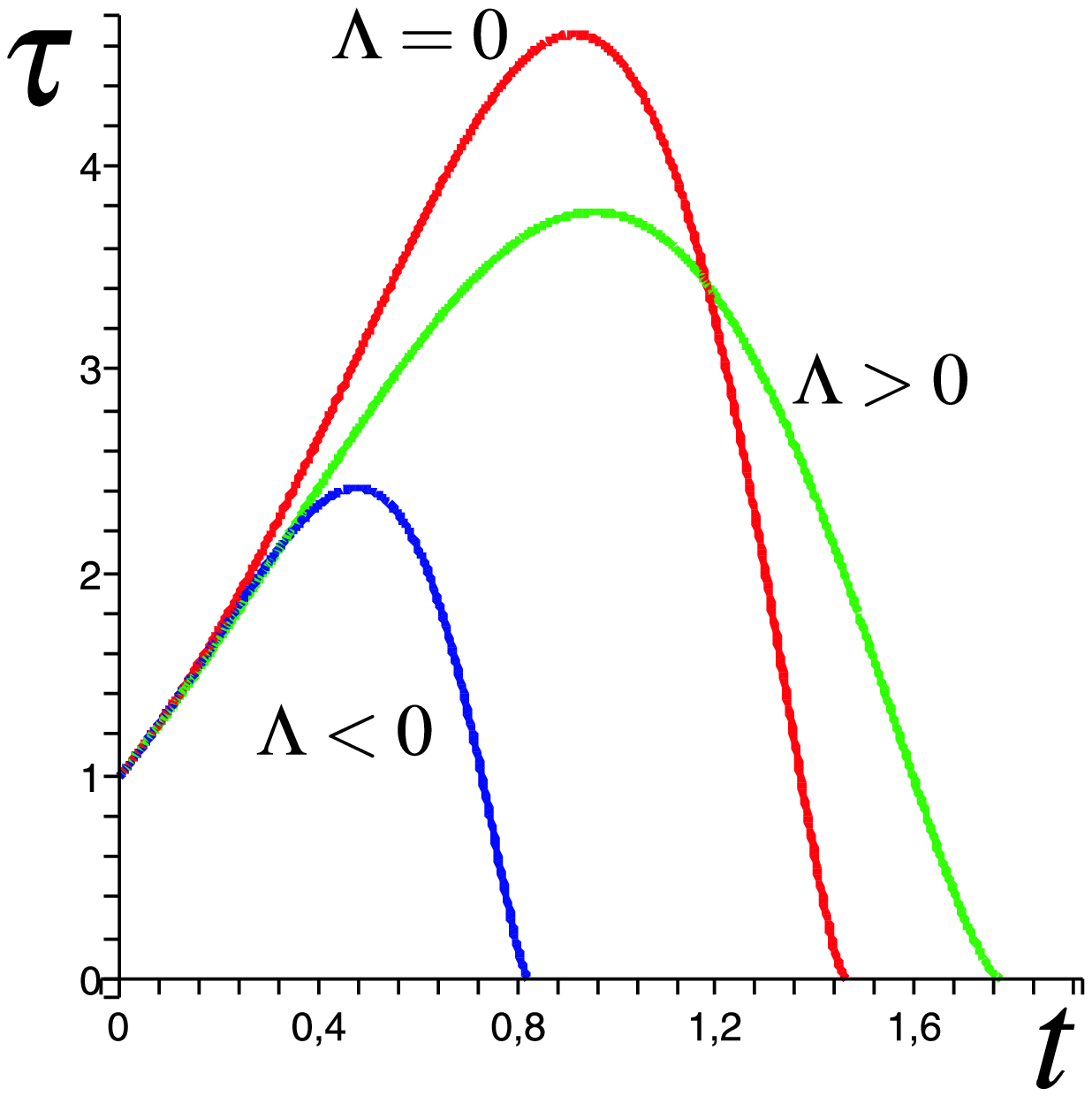}{0.45}{Evolution of the universe with $\delta < 0$ for
different $\Lambda$.}{0.45}{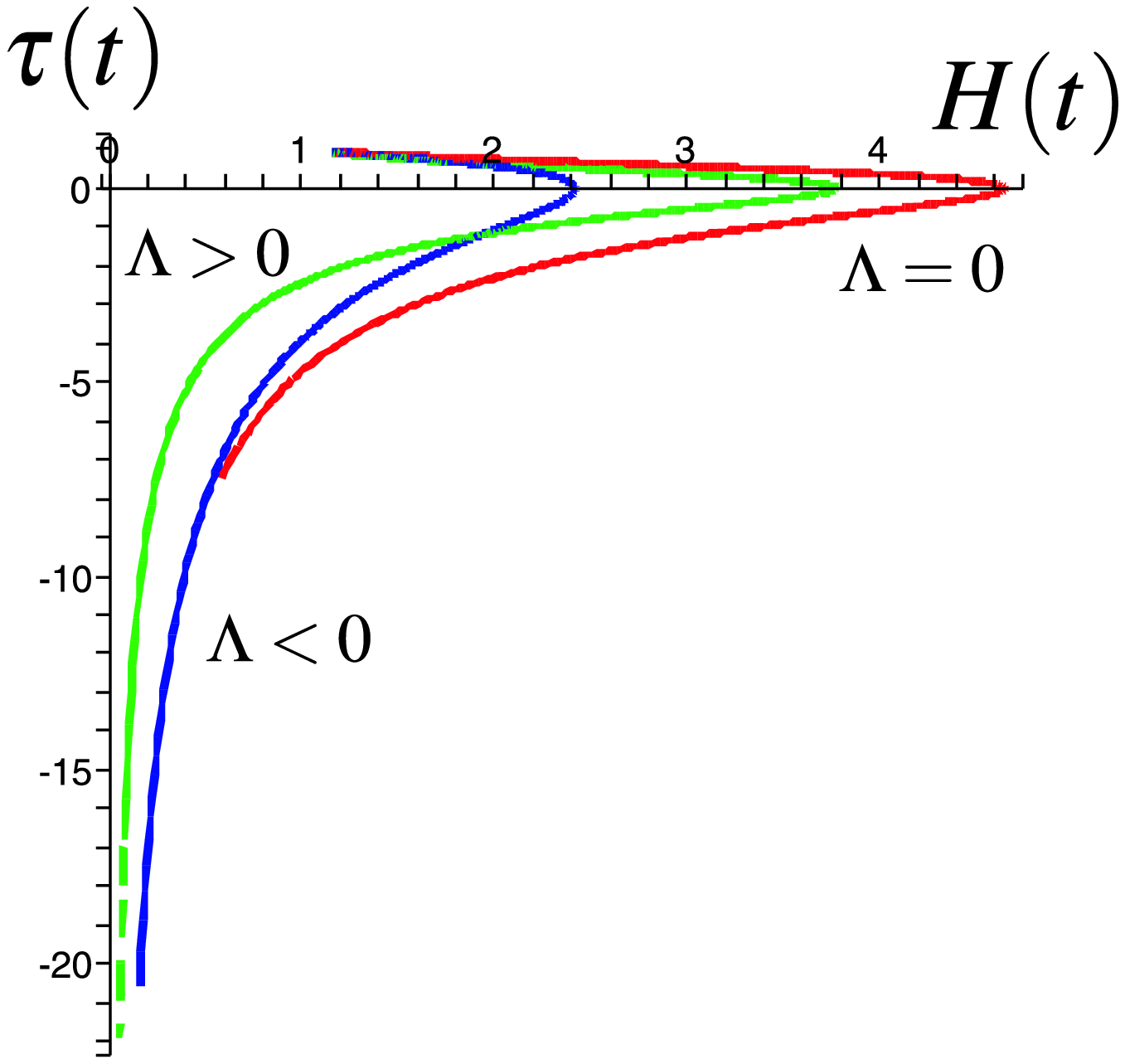}{0.45}{Phase diagram corresponding to
the case with $\delta < 0$ for different $\Lambda$.}{0.45}


               \section{Conclusion}
We consider the self consistent system of spinor, scalar and
gravitational fields within the framework of Bianchi type-I
cosmological model filled with viscous fluid. Solutions to the corresponding
equations are given in terms of the volume scale of the BI space-time, i.e.,
in terms of $\tau = a b c$. The system of equations for
determining $\tau$, energy-density of the viscous fluid $\ve$ and
Hubble parameter $H$ has been worked out. Exact solution to the
aforementioned system has been given only for some special choice of
viscosity. It should be noted that the system \eqref{HVen} is far richer
and allows a number of mathematically interesting results, though not
all of them is physically realizable. Given this fact we plan to review
this system and give a detailed analysis and qualitative solutions of the
corresponding system in some of our future works.


\section{Appendix}

Let us now construct the invariants of spinor field. Since $\psi$ and $\psi^{\star}$ 
(complex conjugate of $\psi$) has $4$ component each, one can construct
$4\cdot 4 = 16$ independent bilinear combinations. They are scalar, pseudoscalar,
vector, axial vector, and tensor denoted, respective, by 
\begin{subequations}
\label{bf}
\begin{eqnarray}
 S&=& \bar \psi \psi,   \\
  P&=& i \bar \psi \gamma^5 \psi, \\
 v^\mu &=& (\bar \psi \gamma^\mu \psi),\\
 A^\mu &=&(\bar \psi \gamma^5 \gamma^\mu \psi), \\
Q^{\mu\nu} &=&(\bar \psi \sigma^{\mu\nu} \psi),
\end{eqnarray}
\end{subequations}
where $\sigma^{\mu\nu}\,=\,(i/2)[\gamma^\mu\gamma^\nu\,-\,
\gamma^\nu\gamma^\mu]$ is the anti-symmetric tensor. Invariants, corresponding to 
the bilinear forms are
\begin{subequations}
\label{invariants}
\begin{eqnarray}
I_S &=& S^2 = (\bar \psi \psi)^2, \\
I_P &=& P^2 = (i \bar \psi \gamma^5 \psi)^2, \\
I_v &=& v_\mu\,v^\mu\,=\,(\bar \psi \gamma^\mu \psi)\,g_{\mu\nu}
(\bar \psi \gamma^\nu \psi),\\
I_A &=& A_\mu\,A^\mu\,=\,(\bar \psi \gamma^5 \gamma^\mu
\psi)\,g_{\mu\nu}
(\bar \psi \gamma^5 \gamma^\nu \psi), \\
I_Q &=& Q_{\mu\nu}\,Q^{\mu\nu}\,=\,(\bar \psi \sigma^{\mu\nu}
\psi)\, g_{\mu\alpha}g_{\nu\beta}(\bar \psi \sigma^{\alpha\beta}
\psi).
\end{eqnarray}
\end{subequations}
$\gamma$ matrices in the above expressions obey the
following algebra
\begin{equation}
\gamma^\mu \gamma^\nu + \gamma^\nu \gamma^\mu = 2 g^{\mu\nu}
\label{al}
\end{equation}
and are connected with the flat space-time Dirac matrices $\bg$ in
the following way
\begin{equation}
 g_{\mu \nu} (x)= e_{\mu}^{a}(x) e_{\nu}^{b}(x) \eta_{ab},
\quad \gamma_\mu(x)= e_{\mu}^{a}(x) \bg_a, \label{dg}
\end{equation}
where $\eta_{ab}= {\rm diag}(1,-1,-1,-1)$ and $e_{\mu}^{a}$ is a
set of tetrad 4-vectors.

For the diagonal metric such as Bianchi-I 
\begin{eqnarray}
ds^2 = dt^2 - a^2(t) dx^2 - b^2(t) dy^2 - c^2(t) dz^2. \nonumber
\end{eqnarray}
we have
\begin{eqnarray}
\gamma_0 &=& \bg_0,\quad \gamma_1 = a(t)\bg_1,\quad
\gamma_2= b(t)\bg_2,\quad \gamma_3 = c(t) \bg_3, \nonumber\\ \\
\gamma^0 &=& \bg^0,\quad \gamma^1 =\bg^1 /a(t),\quad \gamma^2=
\bg^2 /b(t),\quad \gamma^3 = \bg^3 /c(t). \nonumber
\end{eqnarray}
Flat space-time matrices $\bg$ we will choose in the form:
\begin{eqnarray}
\bg^0&=&\left(\begin{array}{cccc}1&0&0&0\\0&1&0&0\\
0&0&-1&0\\0&0&0&-1\end{array}\right), \quad
\bg^1\,=\,\left(\begin{array}{cccc}0&0&0&1\\0&0&1&0\\
0&-1&0&0\\-1&0&0&0\end{array}\right), \nonumber\\
\bg^2&=&\left(\begin{array}{cccc}0&0&0&-i\\0&0&i&0\\
0&i&0&0\\-i&0&0&0\end{array}\right), \quad
\bg^3\,=\,\left(\begin{array}{cccc}0&0&1&0\\0&0&0&-1\\
-1&0&0&0\\0&1&0&0\end{array}\right).  \nonumber
\end{eqnarray}
Defining $\gamma^5$ as follows,
\begin{eqnarray}
\gamma^5&=&-\frac{i}{4} E_{\mu\nu\sigma\rho}\gamma^\mu\gamma^\nu
\gamma^\sigma\gamma^\rho, \quad E_{\mu\nu\sigma\rho}= \sqrt{-g}
\ve_{\mu\nu\sigma\rho}, \quad \ve_{0123}=1,\nonumber \\
\gamma^5&=&-i\sqrt{-g} \gamma^0 \gamma^1 \gamma^2 \gamma^3
\,=\,-i\bg^0\bg^1\bg^2\bg^3 = \bg^5, \nonumber
\end{eqnarray}
we obtain
\begin{eqnarray}
\bg^5&=&\left(\begin{array}{cccc}0&0&-1&0\\0&0&0&-1\\
-1&0&0&0\\0&-1&0&0\end{array}\right).\nonumber
\end{eqnarray}
Note that $\psi$ is a 4 component function given by,
\begin{eqnarray}
\psi =\left(\begin{array}{c}\psi_1\\\psi_2\\
\psi_3\\\psi_4\end{array}\right), \quad \bar \psi = \psi^* \gamma^0 = (\psi_1^*,\, \psi_2^*,\,
-\psi_3^*,\, -\psi_4^*),
\end{eqnarray}
Denoting $\bar F$ bilinear spinor form in Minkowski spacetime and $F$ in 
curve spacetime (in our case in BI) we find the following expressions for the
non-trivial components of bilinear spinor form:
\begin{eqnarray}
\bar S &=& (\psi_1^*\psi_1 + \psi_2^*\psi_2 - \psi_3^*\psi_3 - \psi_4^*\psi_4), \quad S = \bar S,\\
\bar P &=& -i(\psi_1^*\psi_3 + \psi_2^*\psi_4 - \psi_3^*\psi_1 - \psi_4^*\psi_2), \quad P = \bar P,\\
\bar V^0 &=& (\psi_1^*\psi_1 + \psi_2^*\psi_2 + \psi_3^*\psi_3 + \psi_4^*\psi_4), \quad V^0 = \bar V^0,\\
\bar V^1 &=& (\psi_1^*\psi_4 + \psi_2^*\psi_3 + \psi_3^*\psi_2 + \psi_4^*\psi_1), \quad V^1 = \bar V^1/a,\\
\bar V^2 &=& -i(\psi_1^*\psi_4 - \psi_2^*\psi_3 + \psi_3^*\psi_2 - \psi_4^*\psi_1), \quad V^2 = \bar V^2/b,\\
\bar V^3 &=& (\psi_1^*\psi_3 - \psi_2^*\psi_4 + \psi_3^*\psi_1 - \psi_4^*\psi_2), \quad V^3 = \bar V^3/c,\\
\bar A^0 &=& (\psi_1^*\psi_3 + \psi_2^*\psi_4 + \psi_3^*\psi_1 + \psi_4^*\psi_2), \quad A^0 = \bar A^0,\\
\bar A^1 &=& (\psi_1^*\psi_2 + \psi_2^*\psi_1 + \psi_3^*\psi_4 + \psi_4^*\psi_3), \quad A^1 = \bar A^1/a,\\
\bar A^2 &=& -i(\psi_1^*\psi_2 - \psi_2^*\psi_1 + \psi_3^*\psi_4 - \psi_4^*\psi_3), \quad A^2 = \bar A^2/b,\\
\bar A^3 &=& (\psi_1^*\psi_1 - \psi_2^*\psi_2 + \psi_3^*\psi_3 - \psi_4^*\psi_4), \quad A^3 = \bar A^3/c,\\
\bar Q^{01} &=& i(\psi_1^*\psi_4 + \psi_2^*\psi_3 - \psi_3^*\psi_2 - \psi_4^*\psi_1), \quad Q^{01} = \bar Q^{01}/a,\\
\bar Q^{02} &=& (\psi_1^*\psi_4 - \psi_2^*\psi_3 - \psi_3^*\psi_2 + \psi_4^*\psi_1), \quad Q^{02} = \bar Q^{02}/b,\\
\bar Q^{03} &=& i(\psi_1^*\psi_3 - \psi_2^*\psi_4 - \psi_3^*\psi_1 + \psi_4^*\psi_2), \quad Q^{03} = \bar Q^{03}/c,\\
\bar Q^{12} &=& (\psi_1^*\psi_1 - \psi_2^*\psi_2 - \psi_3^*\psi_3 + \psi_4^*\psi_4), \quad Q^{12} = \bar Q^{12}/ab,\\
\bar Q^{23} &=& (\psi_1^*\psi_2 + \psi_2^*\psi_1 - \psi_3^*\psi_4 - \psi_4^*\psi_3), \quad Q^{23} = \bar Q^{23}/bc,\\
\bar Q^{13} &=& i(\psi_1^*\psi_2 - \psi_2^*\psi_1 - \psi_3^*\psi_4 + \psi_4^*\psi_3), \quad Q^{13} = \bar Q^{13}/ac.\\
\end{eqnarray}
Using the above expressions we find
\begin{eqnarray}
I_S &=& S^2 = (\psi_1^*\psi_1)^2 + (\psi_2^*\psi_2)^2 + (\psi_3^*\psi_3)^2 + (\psi_4^*\psi_4)^2 \nonumber\\
&+&2[\psi_1^*\psi_1\psi_2^*\psi_2 - \psi_1^*\psi_1\psi_3^*\psi_3 - \psi_1^*\psi_1\psi_4^*\psi_4 \nonumber\\
    & - & \psi_2^*\psi_2\psi_3^*\psi_3 - \psi_2^*\psi_2\psi_4^*\psi_4 + \psi_3^*\psi_3\psi_4^*\psi_4].   
\end{eqnarray}     

\begin{eqnarray}
I_P &=& P^2 = -(\psi_1^*\psi_3)^2 - (\psi_2^*\psi_4)^2 - (\psi_3^*\psi_1)^2 - (\psi_4^*\psi_2)^2 \nonumber\\
&-&2[\psi_1^*\psi_3\psi_2^*\psi_4 - \psi_1^*\psi_1\psi_3^*\psi_3 - \psi_1^*\psi_3\psi_4^*\psi_2 \nonumber\\
    & - & \psi_2^*\psi_4\psi_3^*\psi_1 - \psi_2^*\psi_2\psi_4^*\psi_4 + \psi_3^*\psi_1\psi_4^*\psi_2].
\end{eqnarray} 

\begin{eqnarray}
I_V &=& (\bar V^0)^2 - (\bar V^1)^2 - (\bar V^2)^2 - (\bar V^3)^2  \nonumber \\
 &=& (\psi_1^*\psi_1)^2 + (\psi_2^*\psi_2)^2 + (\psi_3^*\psi_3)^2 + (\psi_4^*\psi_4)^2 \nonumber\\
& - &(\psi_1^*\psi_3)^2 - (\psi_2^*\psi_4)^2 - (\psi_3^*\psi_1)^2 - (\psi_4^*\psi_2)^2 \nonumber\\
&+&2[\psi_1^*\psi_1\psi_2^*\psi_2 - \psi_1^*\psi_1\psi_4^*\psi_4 - \psi_2^*\psi_2\psi_3^*\psi_3
+ \psi_3^*\psi_3\psi_4^*\psi_4 \nonumber\\ 
&-& \psi_1^*\psi_4\psi_2^*\psi_3
     -\psi_3^*\psi_2\psi_4^*\psi_1 + \psi_1^*\psi_3\psi_4^*\psi_2 + \psi_2^*\psi_4\psi_3^*\psi_1]\nonumber\\
     &=& I_S + I_P.
\end{eqnarray} 

\begin{eqnarray}
I_A &=& (\bar A^0)^2 - (\bar A^1)^2 - (\bar A^2)^2 - (\bar A^3)^2  \nonumber \\
 &=& -(\psi_1^*\psi_1)^2 - (\psi_2^*\psi_2)^2 - (\psi_3^*\psi_3)^2 - (\psi_4^*\psi_4)^2 \nonumber\\
&+ &(\psi_1^*\psi_3)^2 + (\psi_2^*\psi_4)^2 + (\psi_3^*\psi_1)^2 + (\psi_4^*\psi_2)^2 \nonumber\\
&-&2[\psi_1^*\psi_1\psi_2^*\psi_2 - \psi_1^*\psi_1\psi_4^*\psi_4 - \psi_2^*\psi_2\psi_3^*\psi_3
+ \psi_3^*\psi_3\psi_4^*\psi_4 \nonumber\\ &-& \psi_1^*\psi_4\psi_2^*\psi_3
     -\psi_3^*\psi_2\psi_4^*\psi_1 + \psi_1^*\psi_3\psi_4^*\psi_2 + \psi_2^*\psi_4\psi_3^*\psi_1]\nonumber\\
     &=& -(I_S + I_P).
\end{eqnarray}

\begin{eqnarray}
I_Q &=& 2[(\bar Q^{12})^2 + (\bar Q^{13})^2 + (\bar Q^{23})^2 - (\bar Q^{01})^2 
-  (\bar Q^{02})^2 - (\bar Q^{03})^2] \nonumber \\
 &=& (\psi_1^*\psi_1)^2 + (\psi_2^*\psi_2)^2 + (\psi_3^*\psi_3)^2 + (\psi_4^*\psi_4)^2 \nonumber\\
&+ &(\psi_1^*\psi_3)^2 + (\psi_2^*\psi_4)^2 + (\psi_3^*\psi_1)^2 + (\psi_4^*\psi_2)^2 \nonumber\\
&+&2[\psi_1^*\psi_1\psi_2^*\psi_2 - \psi_1^*\psi_1\psi_4^*\psi_4 - \psi_2^*\psi_2\psi_3^*\psi_3
+ \psi_3^*\psi_3\psi_4^*\psi_4 \nonumber\\ &+& \psi_1^*\psi_4\psi_2^*\psi_3
     +\psi_3^*\psi_2\psi_4^*\psi_1 - \psi_1^*\psi_3\psi_4^*\psi_2 - \psi_2^*\psi_4\psi_3^*\psi_1]\nonumber\\
    & - & 4 (\psi_1^*\psi_1\psi_3^*\psi_3 +  \psi_2^*\psi_2\psi_4^*\psi_4) = 2(I_S - I_P).
\end{eqnarray} 

Thus we see that that the invariants $I_V$, $I_A$ and $I_Q$ can be expressed in terms of $I_S$ and $I_P$.

An alternative proof of $I_V = - I_A = (I_S + I_P)$ and $I_Q = 2 (I_S - I_P)$ can be given 
using Feirz transformation. In doing so we recall that any $4x4$ matrix $\G$ can be presented as a linear
combination of $\gamma^A$:
\begin{equation}
\G = \sum_A c_A \gamma^A, \quad c_A = \frac{1}{4} Tr \gamma_A \G.
\end{equation}
Explicitly it can be presented as
\begin{equation}
\G_{ik} = \frac{1}{4} \sum_A \G_{lm}\gamma^A_{ml}\gamma_{Aik}.
\end{equation}
Here
\begin{equation}
\gamma^A = \{I,\, \gamma^5,\,
\gamma^\mu,\,i\gamma^\mu\gamma^5,\,i\sigma^{\mu,\nu}\},\quad
A=1,2..16.
\end{equation}
The completeness condition gives
\begin{equation}
\delta_{ij}\delta_{kl} = \frac{1}{4} \sum_A \gamma_{Aik}
\gamma^A_{kl}.
\end{equation}
Multiplying the foregoing expression by $\bp^a_i \psi^b_k
\bp^c_m\psi^d_l$ one gets
\begin{equation}
(\bp^a\psi^d)(\bp^c\psi^b) = \frac{1}{4} \sum_A (\bp^a \gamma_A
\psi^b)(\bp^c \gamma^A \psi^d).
\end{equation}
Replacing $\psi^d \to \gamma^B \psi^d$ and $\psi^b \to \gamma^C
\psi^b$ and using the expression
\begin{equation}
\gamma^A \gamma^B = \sum_R C_R \gamma^R,\quad C_R = \frac{1}{4} Tr
\gamma^A \gamma^B \gamma_R,
\end{equation}
one gets the other identities. Denoting $I_S =
(\bp^a\psi^b)(\bp^c\psi^d), \quad I_S^\prime =
(\bp^a\psi^d)(\bp^c\psi^b) \cdots...$ one finds \cite{fierz}
\begin{eqnarray}
4 I_S^\prime & = & I_S + I_P + I_V + I_Q + I_A,\\
4 I_P^\prime & = & I_S + I_P - I_V + I_Q - I_A,\\
4 I_V^\prime & = & 4 I_S - 4 I_P - 2 I_V + 2 I_A,\\
4 I_Q^\prime & = & 6 I_S + 6 I_P - 2 I_Q,\\
4 I_A^\prime & = & 4 I_S - 4 I_P + 2 I_V - 2 I_A.
\end{eqnarray}
After a little manipulations from the foregoing system one
comes to desired relations between the invariants of the bilinear 
spinor fields.

\newcommand{\hnl}{\htmladdnormallink}

\end{document}